\documentclass{article}
\usepackage[utf8]{inputenc}
\usepackage{spconf,amsmath,graphicx,hyperref}
\usepackage{amsmath}
\usepackage{amssymb} 

\title{Audio-Visual Speech Enhancement in Complex Scenarios with Separation and Dereverbration Joint Modeling}
\name{
    Jiarong Du$^{1,2}$, Zhan Jin$^{3,4}$, Peijun Yang$^{1,2}$, Juan Liu$^{2,1*}$\thanks{*Corrsponding author: liujuan@whu.edu.cn.} , Zhuo Li$^{5}$, Xin Liu$^{5}$, Ming Li$^{2,4*}$\thanks{*Corrsponding author: ming.li369@dukekunshan.edu.cn; }
}
\address{
    $^1$School of Cyber Science and Engineering, Wuhan University, Wuhan, China\\
    $^2$School of Artificial Intelligence, Wuhan University, Wuhan, China\\
    $^3$School of Computer Science, Wuhan University, Wuhan, China\\
    $^4$Suzhou Municipal Key Laboratory of Multimodal Intelligent Systems,\\ Digital Innovation Research Center, Duke Kunshan University, Kunshan, China\\
    $^5$Hardware Engineering System, OPPO, Beijing, China\\
}
\begin{document}
%
\maketitle
\begin{abstract}
Audio-visual speech enhancement(AVSE) is a task that uses visual auxiliary information to extract a target speaker’s speech from mixed audio. In real-world scenarios, there often exist complex acoustic environments, accompanied by various interfering sounds and reverberation. Most previous methods struggle to cope with such complex conditions, resulting in poor perceptual quality of the extracted speech. In this paper, we propose an effective AVSE system that performs well in complex acoustic environments. Specifically, we design a ``separation before dereverberation'' pipeline that can be extended to other AVSE networks. The 4th COG-MHEAR Audio-Visual Speech Enhancement Challenge (AVSEC) aims to explore new approaches to speech processing in multimodal complex environments. We validated the performance of our system in AVSEC-4: we achieved excellent results in the three objective metrics on the competition leaderboard, and ultimately secured first place in the human subjective listening test. Our demos are available at \footnote{\url{https://jarendd.github.io/}}.

\end{abstract}

\section{Introduction}
\label{sec:intro}

As described by the classic cocktail-party effect \cite{arons1992review}, humans can always focus on the sounds they want to hear in noisy environments. Blind Source Separation (BSS) methods \cite{luo2019conv,wang2023tf} rely on prior information regarding the number of speakers—a constraint that limits their applicability in many practical scenarios. In contrast, Target Speaker Extraction (TSE), with the help of prompt information, can focus on selecting the desired speech from mixed audio. Previous TSE methods \cite{xu2020spex,ge2020spex+,hao2024x} often used the target speaker’s recording as registration information. However, high-quality registration information is difficult to obtain in advance. 

With the development of multimodal technology \cite{michelsanti2021overview}, researchers have gradually shifted their focus from audio-only methods to audio-visual and multimodal methods. The role of multimodal information, such as registered speech, directional cues, and lip movements, was explored by Gu et al. \cite{gu2020multi}. Meanwhile, models designed for lip-reading tasks \cite{martinez2020lipreading} have demonstrated a remarkable capability to capture the correlation between lip movements and speech. Building on this foundation, Wu et al. \cite{wu2019time} utilized a pre-trained lip-reading model to extract the speaker's lip movement features, which were subsequently fused with speech features within their network. Subsequent work by Pan et al. \cite{pan2021muse,pan2022usev} conducted extensive studies on the fusion methods between visual features and mixed audio. Additionally, \cite{li2024audio,li2024iianet} were inspired by the real structure and operational processes of the human brain, and verified that appropriate data fusion methods can enhance the performance of multimodal TSE  models. Although these models have shown excellent performance in simulated datasets with the speech mixture setting, most studies fail to account for complex acoustic conditions in real-world scenarios—such as reverberation and diverse strong noise sources. In this case, models trained with data simulated by simple mixing may degrade the quality of extracted speech; while training directly with complex data simulated with real Room Impulse Responses (RIR) may also make the model difficult to generate the clean and dereverberated speech of the target speaker.

To address the issues, this paper proposes several improvements based on \cite{jin2024target}, enabling the extracted speech from complex environments to maintain good perceptual quality. First, to avoid training the model directly on complex mixed data, we designed a more reasonable dynamic data mixing method, laying the foundation for subsequent progressive training. Second, targeting at the increased difficulties of model learning caused by complex training data, this paper proposes a two-stage ``separation before dereverberation'' approach. Specifically, progressive loss training is incorporated into the separation network in the first stage to optimize the learning process; for the reverberant speech obtained after separation, a pre-trained dereverberation model is employed to predict the clean speech as a post-processing module. Moreover, after the competition, we adopted a joint training strategy to further improve the model's performance. Our system achieved the excellent performance in the three objective metrics—Perceptual Evaluation of Speech Quality (PESQ) \cite{rix2001perceptual}, Short-Time Objective Intelligibility (STOI)~\cite{taal2010short}, Scale-Invariant Signal-to-Distortion Ratio (SISDR) \cite{le2019sdr}—and ranked as the first place in the human subjective evaluation.

\vspace{-10pt}
\section{Methods}
\label{sec:Methods}
\subsection{Overall Model Architecture}
\label{ssec:subhead}
\begin{figure}[htb]
\begin{minipage}[b]{1.0\linewidth}
  \centering
  \centerline{\includegraphics[width=8.5cm]{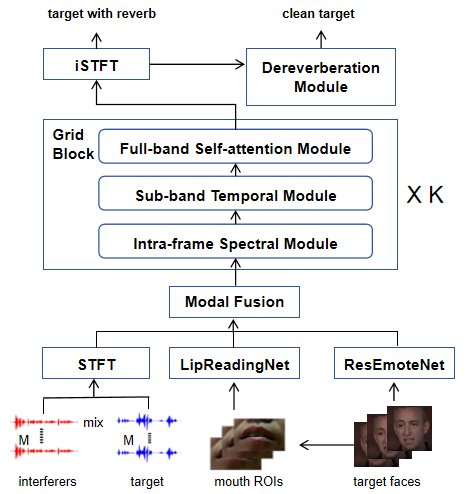}}
\end{minipage}
\caption{Overall Model Architecture. Lip movement and facial expression features are extracted by pre-trained lip-reading model \cite{martinez2020lipreading} and expression estimation model \cite{Roy2024ResEmoteNetBA}, respectively. These three types of features are concatenated along the channel dimension and then fed into the separation module following the same way as in \cite{jin2025Robust}. The separation module adopts the same structure as that of \cite{wang2023tf}.\vspace{-10pt}}
\label{fig:model_arch}
\end{figure}

As a well-established blind separation network, TFGridNet \cite{wang2023tf} includes a separation module, GridBlock, with proven effectiveness. AV-GridNet \cite{pan2023scenario} was the first to integrate video information with TFGridNet for the AVSE task. Building on this, Jin et al. \cite{jin2024target} fuse audio and visual features through a single channel-wise concatenation operation prior to the GridBlocks, reducing the model complexity while achieving State-Of-The-Art (SOTA) results in the 3rd AVSEC. Furthermore, according to research in \cite{jin2025Robust}, in addition to lip movements, facial expressions contain rich information that can serve as new auxiliary information to enhance the performance of AVSE. The overall structure of the network is therefore shown in Figure \ref{fig:model_arch}.

\vspace{-10pt}  

\subsection{Separation Before Dereverberation and Progressive Training}
\label{ssec:SFTD and Progressive Training}

Studies in \cite{zhao22b_interspeech} show that model training with data from  different reverberation conditions can significantly affect the performance. In reverberant scenarios, adopting near-field clean speech as the target increase the training difficulty of the model and reduce the quality of the extracted speech. Using reverberant target can enhance separation but introduce reverberation into the output signal. To tackle the problem, we proposed a two-stage ``separation before dereverberation'' approach. This method decouples the speech extraction and dereverberation processes, enabling the model to ignore reverberation during the separation stage which use the target speaker's speech with reverberation as the target and then perform dereverberation on the separated signal.

\begin{figure}[htb]
\begin{minipage}[b]{1.0\linewidth}
  \centering
  \centerline{\includegraphics[width=8.5cm]{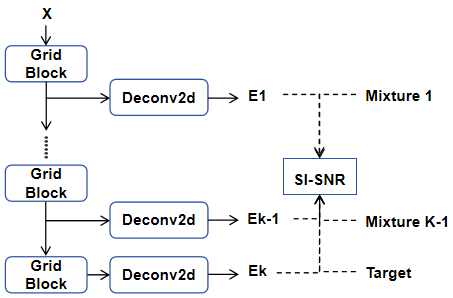}}
\end{minipage}
\caption{Progressive Training. In this work, K is set to 6. $\text{E}_k$ represents the intermediate estimate output by the $k$-th separation module layer. $\text{Mixture}_k$ denotes the intermediate target at the $k$-th layer. Target refers to the clean target speaker's speech with reverberation. \vspace{-10pt}}
\label{fig:progressive_train}
\end{figure}

We employed progressive loss training \cite{hou2024sir} in the separation network of the first stage to make the training process more smooth. During training, the reverberant target and interfering speech are initially mixed at a random Signal-to-Distortion Ratio (SDR) within the range of [-18, 6] dB. Based on this initial SNR, we generate K-1 intermediate targets with a layer-wise increasing SNR. Each intermediate target serves as the ground truth for the output of the corresponding GridBlock, with a 5 dB difference between successive targets. And the final layer directly employs the reverberant target speech as the training target without any additional mixing. As shown in Figure \ref{fig:progressive_train}, after each separation module, we use an independent Deconv2d layer to calculate the intermediate estimates, and compute the SISDR loss \cite{le2019sdr} between each intermediate estimate and its corresponding intermediate target. The total loss function is defined as: 
$$\mathcal{L}_\text{total} = \frac{1}{K} \sum_{k=1}^{K} \mathcal{L}^{(k)}$$ 
where \(K=6\) denotes the number of GridBlocks, \(\mathcal{L}^{(k)}\) represents the loss at the k-th layer, and \(\mathcal{L}_\text{total}\) denotes the total loss of the separation model. This method enables the model to suppress interfering speech in shallow network layers and learn the details of the target speech in deeper layers.

For the extracted speech with reverberation, we use a pre-trained dereverberation model \cite{richter2023speech} for post-processing to obtain the clean speech. Richter et al. \cite{richter2023speech} proposed a score-based generative diffusion model, SGMSE+, which employs a complex-valued Short-Time Fourier Transform (STFT) representation for speech signals. It designs a linear Stochastic Differential Equation (SDE)-based forward diffusion process with a drift term and a score model-based reverse recovery process. Notably, this model only requires 30 diffusion steps to generate high-quality speech. In dereverberation tasks, SGMSE+ significantly outperforms baseline methods in relevant metrics, demonstrating excellent dereverberation performance.

\vspace{-10pt}  
\subsection{In-depth Study of Separation Before Dereverberation Pipeline }
\label{ssec:subhead}
To achieve better dereverberation performance, we chose to use a relatively advanced dereverberation model \cite{richter2023speech} during the competition. As it is a generative model, its training and inference speeds are relatively slow, making it unsuitable for joint training with the separation model. 

Therefore, after the competition, we aimed to extend the post-processing approach into a joint method. To enable joint training with the separation model, we selected SkipConvNet \cite{kothapally2020skipconvnet} as the new dereverberation module. It adopts a relatively simple stacked convolution structure while achieving favorable performance, making it a widely used dereverberation network. To investigate the interaction between the separation and dereverberation modules, we conducted joint training and defined the new loss function as:
$$\mathcal{L}_\text{joint} = \mathcal{L}_\text{separate} + \mathcal{L}_\text{dereverb}$$ 
\(\mathcal{L}_\text{separate}\) represents the progressive loss mentioned in Section \ref{ssec:SFTD and Progressive Training}, while \(\mathcal{L}_\text{dereverb}\) depends specifically on the calculation method of the dereverberation module used. For example, SkipConvNet \cite{kothapally2020skipconvnet} employs the mean squared error between the predicted and target speech spectrograms as its loss function. Training both modules from scratch may lead to performance degradation due to inconsistent optimization directions. However, using pretrained weights for both modules may reduce the gain of joint training. Therefore, we first initialize the model with the weights from the pretrained dereverberation module, and then perform joint training of both the separation and dereverberation modules.

\vspace{-10pt}  
\section{Experiment al RESULTS}
\label{sec:Experiments}

\subsection{Dataset}
\label{ssec:Dataset}
To promote the performance of models' robustness in real-world scenarios, the 4th COG-MHEAR AVSEC \footnote{\url{https://challenge.cogmhear.org/}} designed a complex dataset that is closer to real-world environments. Unlike LRS3 \cite{afouras2018lrs3}, and VOXCELEB2 \cite{chung2018voxceleb2}, the AVSEC-4 dataset includes reverberation simulated from real rooms, and the total number of speakers in each audio clip may exceed two. Furthermore, music noise is added to the data, increasing the complexity of the separation task.

The training set contains 34,524 utterances with a total duration of 113 hours and 17 minutes. The development set includes 3,306 utterances, totaling 8 hours and 38 minutes. The test set comprises 3,180 utterances, of which 1,500 are used for leaderboard evaluation and the rest for listening tests. The test set is produced in real conference room scenarios by controlling the room size and sound propagation distance, with the SNR ranging from -18 dB to 6.55 dB.

\vspace{-10pt}  
\subsection{Implementation Details}
\label{ssec:Implementation details}

We dynamically mixed the binaural target signals and interferers at SNR ranging from -18 dB to 6 dB. The mixed audio segments and corresponding target speaker videos (sampled at 25 frames per second) were randomly truncated into 3-second chunks. We used the Adam optimizer with an initial learning rate of 0.001. The learning rate was halved when the best validation loss showed no improvement for 3 consecutive epochs, and training was stopped when there was no improvement for 10 consecutive epochs. The SISDR loss \cite{le2019sdr} was adopted as the training loss function, with the batch size fixed at 2.

\begin{table}[htbp]
\centering 
\renewcommand{\arraystretch}{1} 
\caption{The leaderboard of AVSEC-4. All results evaluated on 1,500 utterances in the test set. M denotes monaural audio, while B denotes binaural audio.}\label{tab1}
\begin{tabular}{|c|c|c|c|c|} 
\hline
System &Channel & PESQ & STOI & SISDR  \\
\hline
Noisy &M & 1.29 & 0.51 & -25.94  \\
Noisy &B & 1.31 & 0.55 & -24.28   \\
\hline
Team-OPTIMAL &M & 	1.37 &	0.48 &	-21.44\\
USTC\_Entry1 &M & 1.35 &	0.52	& -23.96 \\
GU-ENU &M & 1.35&	0.52	& -23.93  \\
Rahma\_Team &M & 1.30 &	0.55	& -24.72  \\
TeamKCW &M & 	1.58 &	0.60	 & -21.77  \\
SND\_VD &M & 1.72&	0.64&	-21.96  \\
R-test1 &B & 1.64	& 0.66	& -19.26 \\
BioASP &M & 1.71	&0.67&	-20.41  \\
SUSTechAILab &B & 1.95 &	0.67	&-18.75  \\
CITISIN &M & \textbf{2.29} & 0.78 & \textbf{-17.15}   \\
\hline
WHU\_DKU(ours) &B & 2.16  & \textbf{0.82} & -17.47\\
\hline
\end{tabular}
\vspace{-10pt}  
\end{table}

\begin{table}[htbp]
\centering 
\setlength{\tabcolsep}{10pt} 
\renewcommand{\arraystretch}{1} 
\caption{Evaluation results on the complete test set.}\label{tab2}
\begin{tabular}{|c|c|c|c|} 
\hline
System & PESQ & STOI & SISDR \\
\hline
Noisy & 1.08 & 0.55 & -23.93 \\
\hline
WHU\_DKU(ours)& \textbf{1.60}  & \textbf{0.83} & \textbf{-17.16} \\
\hline
\end{tabular}
\vspace{-10pt}
\end{table}

In addition, after the competition, We investigated the impact of monaural output on system performance and explored a different training strategy. During the training, the number of epochs for learning rate reduction and early stopping was adjusted from 3 and 10 to 15 and 30, respectively. Once the learning rate dropped to a certain level, we added the STFT loss \cite{pan2022hybrid} function to the original SISDR loss \cite{le2019sdr} function. This enables the model to gain spectral benefits, and enhance the listening quality of the extracted speech.

\vspace{-10pt}  
\subsection{Results and Discussions}
\label{ssec:subhead}
The detailed results are shown in Table \ref{tab1}. As can be seen from the first row of the table, the objective metrics of the original mixed data (Noisy) are very low, indicating that the original data is complex and severely degraded by noise and reverberation. The middle part of the table corresponds to other participating teams. Our proposed system (WHU\_DKU) achieved competitive performance across all metrics. With a PESQ of 2.16, we rank second in this indicator, yet our STOI metric reaches 0.82. We believe the high STOI but relatively lower PESQ stems from our system’s strong interference suppression. However, this aggressive suppression can overoptimize noise removal, causing subtle damage to the natural timbre of the original speech.

\begin{figure}[htb]
\begin{minipage}[b]{1.0\linewidth}  
  \centering
  \includegraphics[width=\linewidth]{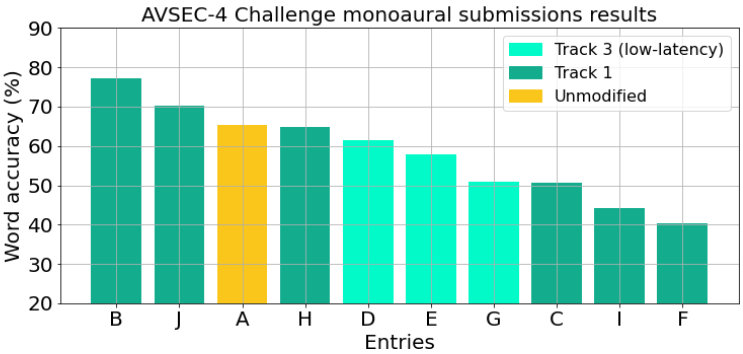}  
\end{minipage}

\begin{minipage}[b]{1.0\linewidth}  
  \centering
  \includegraphics[width=\linewidth]{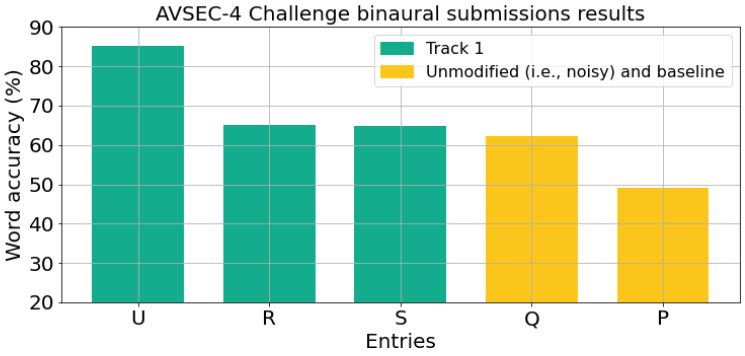}
\end{minipage}

\caption{Results of subjective evaluation by human auditory perception. Our system is represented by ``U''.}
\label{fig:subjective_eval}
\vspace{-10pt}
\end{figure}

In the human-ear subjective evaluation, participants viewed 126 video clips, 120 samples, and 6 validation questions. The performance of each system was evaluated by calculating the word intelligibility score, where a higher score indicates better intelligibility. The final score was based on the average of all participants' ratings. The evaluation results are presented in Figure \ref{fig:subjective_eval}, and our system is represented by system ``U''. Our score was 85.23, the only one that exceeds 80 points, which demonstrates the effectiveness of our system. Although our system does not have a significant advantage in the objective metrics, its leading performance in subjective ratings demonstrates that our method is beneficial for human auditory perception.

After the competition, we re-evaluated the system's performance on the complete monoaural test set. We employed the second training strategy mentioned in Section \ref{ssec:Implementation details}, and output monaural audio. As shown in Table \ref{tab2}, significant improvements were observed across all three objective metrics. However, due to its prohibitively long training time, this strategy was not adopted for subsequent experiments.

\begin{table}[htbp]
\centering 
\setlength{\tabcolsep}{6pt} 
\renewcommand{\arraystretch}{1} 
\caption{Evaluation results on the complete test set.}\label{tab3}
\begin{tabular}{|c|c|c|c|} 
\hline
System & PESQ & STOI & SISDR \\
\hline
Derev before Sep & 1.35 & 0.69 & -23.71 \\
\hline
Separation only& 1.42  & 0.78& -18.22 \\
Sep before Derev& 1.44  & 0.81& -18.71 \\
Separation only (joint)& 1.44  & 0.79 & -18.07 \\
Sep before Derev (joint) & \textbf{1.69}  & \textbf{0.83} & \textbf{-17.66} \\
\hline
\end{tabular}
\vspace{-10pt}
\end{table}

Moreover, ablation studies of different training strategies are shown in Table \ref{tab3}. The approach of "dereverberation before separation" achieves the worst performance. The dereverberation operation will change the characteristics of the original mixed audio, making the subsequent separation network difficult to perform well. In contrast, ``separation before dereverberation'' can effectively combine the strength of the two tasks, achieving better results. And joint training can further enhance the quality of the speech. Furthermore, the joint training of the two modules also have a positive impact on the separation network using alone, resulting in a certain improvement in the separation results of the first stage.

\section{Conclusion}
\label{sec:Conclusion}

To address the limitations of existing AVSE methods in complex acoustic environments, this paper introduces an effective AVSE system along with a corresponding ``separation before dereverberation'' pipeline. The proposed approach reduces the learning complexity associated with processing complex data, improves the quality of the enhanced speech, and achieves SOTA performance in the AVSEC-4.



\vfill\pagebreak


\begingroup
  \small  
  \bibliographystyle{IEEEbib}
  \bibliography{strings,refs}
\endgroup

\end{document}